\documentclass[aps, prd, twocolumn, superscriptaddress, nofootinbib]{revtex4}
\usepackage{dcolumn}
\usepackage{amssymb}
\usepackage{amsmath,amssymb,amsfonts}

\begin{document}

\title{Intermediate inflation from rainbow gravity}
\author{John D. Barrow}
\affiliation{DAMTP, Centre for Mathematical Sciences, University of Cambridge,
Wilberforce Rd., Cambridge, CB3 0WA,UK}
\author{Jo\~{a}o Magueijo}
\affiliation{Theoretical Physics, Blackett Laboratory, Imperial College, London, SW7 2BZ,
United Kingdom}
\affiliation{Dipartimento di Fisica, Universita La Sapienza and Sezione Roma1 INFN, P.le A. Moro 2, 00185 Roma, Italia}
\date{\today }

\begin{abstract}
It is possible to dualize theories based on deformed dispersion relations
and Einstein gravity so as to map them into theories with trivial dispersion
relations and rainbow gravity. This often leads to \textquotedblleft dual
inflation\textquotedblright\ without the usual breaking of the strong energy
condition. We identify the dispersion relations in the original frame which
map into \textquotedblleft intermediate\textquotedblright\ inflationary
models. These turn out to be particularly simple: power-laws modulated by 
powers of a logarithm. 
The fluctuations predicted by these scenarios are near, but not
exactly scale-invariant, with a red running spectral index. 
These dispersion relations deserve further study within the
context of quantum gravity and the phenomenon of dimensional reduction in
the ultraviolet.
\end{abstract}

\maketitle


\section{Introduction}

In recent work~\cite{dimred,rainbowred,measure} we have investigated the
viability of cosmological scenarios based on modified dispersion relations
(MDRs), when combined with Einstein gravity. Notably, it was found that the
MDR 
\begin{equation}
E^{2}=c^{2}p^{2}=p^{2}(1+(\lambda p)^{2\gamma }),  \label{ddr1}
\end{equation}%
linking the energy $E$ and momentum $p$, with constant $\gamma $ and
characteristic running scale $\lambda $, is associated with exactly
scale-invariant fluctuations when the constant parameter $\gamma =2$, with
suitable modifications leading to small deviations from exact
scale-invariance~\cite{dimred}. This is particularly interesting, since
these dispersion relations appear to model well the phenomenon of
dimensional reduction in the ultra-violet (UV), for which there is growing
evidence in numerous approaches to quantum gravity~\cite%
{Lollprl,Litim,Reuter,Hl,HLspec,Alesci:2011cg,Benedetti:2008gu,Modesto1,Caravelli,Magliaro,Modesto2,Calcagni1,Calcagni2,visser}%
. The mechanism producing fluctuations is analogous to that of varying speed
of light/sound models~\cite{csdot,bim,mag} and, at face-value, dispenses
with the need for inflation to perform this role.

Nonetheless, it is noteworthy that it is possible to change units, or
\textquotedblleft frame\textquotedblright , so as to render the dispersion
relations trivial. The non-trivial phenomenology of the theory is then
shifted elsewhere, specifically to the theory of gravity. This operation was
performed in~\cite{rainbowred,measure} and it is equivalent to what is done
with Brans-Dicke theory when one conformally changes from the Einstein
frame, with a varying $G$, to the Jordan frame, with a constant $G$ but
modified gravity. Specifically in~\cite{rainbowred} we showed that the new
frame, where the speed of light, $c,$ is constant, is the frame of
\textquotedblleft rainbow gravity\textquotedblright ~\cite{rainbowDSR}, and
that typically one has inflation in this frame, even if the strong energy
conditions are not broken. A similar phenomenon was found before in~\cite%
{Garattini}.

This is by no means always the case. Indeed, the very topical case $\gamma
=2 $ (associated with UV spectral dimension 2) does not lead to
inflation in the dual frame from the point of view of fluctuations; instead,
it leads to the switching off of gravity altogether. Also, whenever we do
have inflation in the dual frame, it is not standard inflation. It is
inflation driven by the gravity theory, rather than by the matter content
(or an \textquotedblleft inflaton field\textquotedblright ). Also, (near)
scale-invariant fluctuations can be obtained under very different conditions
to standard inflation: for example, we do not need
to be near de Sitter. For this reason in~\cite{rainbowred} these models of
inflation were labelled \textquotedblleft esoteric
inflation\textquotedblright .

The specific models derived from (\ref{ddr1}) lead to power-law or de Sitter
inflation. The purpose of this paper is to obtain more general MDRs associated
with intermediate inflation~\cite{intinfl0,intinfl1,intinfl2,intinfl3}. This
form of inflation is a generalisation of de Sitter inflation in which the
expansion scale factor evolves with 
\begin{equation}
a(t)=\exp \{At^{n}\}  \label{int}
\end{equation}
with $A>0$ and $0<n\leq 1$ constants. Subject to Einstein gravity 
it creates scale-invariant fluctuations
when $n=2/3$ as well as $n=1$ (which is de Sitter). It has been found to
arise in a wide class of scalar-tensor gravity theories \cite{intinfl4} 
and in general relativistic
cosmologies where there is an effective equation of state, linking the
density $\rho $ and the pressure $P,$ of the form $\rho +P=\Gamma \rho ^{B}$%
. For $\Gamma \neq 0$ and $B\neq \frac{1}{2}$ or $1$, a zero-curvature FRW
universe has an exact solution of the form (\ref{int}) with \cite{intinfl3}
\begin{eqnarray}
n&=&\frac{2(1-B)}{1-2B}\\
A&=&\frac{3^{B/(1-2B)}\Gamma ^{1/(1-2B)}(B-%
\frac{1}{2})^{2(1-B)/(1-2B)}}{B-1}.  \label{int2}
\end{eqnarray}%
This is equivalent to a family of exact solutions containing a single scalar
field $\phi $ with a particular self-interaction potential, $V(\phi )$ \cite%
{intinfl3}.

\section{Rainbow inflation revisited}

\label{rainbinfl} It was proved in~\cite{rainbowred} that MDRs of the form (%
\ref{ddr1}) combined with Einstein gravity can be mapped into a rainbow
frame with trivial dispersion relations but a modified theory of gravity. In
general, this modified gravity theory is very different, but it was shown in~%
\cite{rainbowred} that for \textit{background solutions} with no curvature
(i.e. FRW models with $K=0$) this amounts to keeping Einstein gravity and
adopting an \textquotedblleft effective\textquotedblright\ equation of state
in the rainbow frame with 
\begin{equation}
{\tilde{w}}=w-\frac{2}{3}\gamma ,  \label{neww}
\end{equation}%
where $w=P/\rho $ is the linear equation of state factor in the Einstein
frame. We stress that the modified gravity theory is more complex in
general, in particular for the perturbations around these solutions. Here we
present an alternative derivation of this result which is not only
particularly simple, but will mimic the method used for finding intermediate
inflationary solutions in the next Section.

If (\ref{ddr1}) is valid then at high energies (in the \textquotedblleft
UV\textquotedblright\ limit) we have: 
\begin{equation}
c\propto (\lambda p)^{\gamma }\,.
\end{equation}%
Here $p$ is the physical momentum, so if we focus on a comoving mode
labelled by $k$, then: 
\begin{equation}
p=\frac{k}{a},
\end{equation}%
and this property is valid so long as there is spatial translational
invariance. Therefore, in the Einstein frame, $c$ is both energy and time
dependent, due to the expansion. We may define the rainbow frame (in which $%
c $ is constant) directly in terms of proper time, in a procedure that is
equivalent to that used in~\cite{rainbowred}. First define a disformal
transformation by keeping the spatial coordinates and $a$ unchanged but
replacing $t$ by 
\begin{equation}
\tilde{t}=\int dt\,c.  \label{tildet}
\end{equation}%
Since in Einstein gravity for $K=0$ Friedmann expansion we have, 
\begin{equation}
a\propto t^{\frac{2}{3(1+w)}},  \label{aoft}
\end{equation}%
we must also have 
\begin{equation}
\tilde{t}\propto (\lambda k)^{\gamma }t^{1-\frac{2\gamma }{3(1+w)}},
\end{equation}%
so that 
\begin{equation}
a\propto {\left( \frac{\tilde{t}}{(\lambda k)^{\gamma }}\right) }^{\frac{2}{%
3(1+w)-2\gamma }}\;.
\end{equation}%
By comparing with (\ref{aoft}) we can read off that \textit{in this context}
(i.e. for FRW, $K=0$ solutions) the new gravity theory is equivalent to
Einstein gravity, but with the matter content modified according to (\ref%
{neww}). Notice, however, that the Hubble constant is now $k$-dependent,
something that will affect indirectly the perturbations (in addition to the
direct effects of modified gravity).

It turns out that this is the most general situation for which:
\begin{itemize}
\item The speed of light has a power law in the momentum (or energy) in the
Einstein frame.

\item The effective equation of state is a constant in the rainbow frame.
\end{itemize}
Specifically, we have power-law inflation in the rainbow frame if 
\begin{equation}
\frac{1+3w}{2}<\gamma <\frac{3}{2}(1+w)
\end{equation}%
and de Sitter inflation if we saturate the second identity: 
\begin{equation}
\gamma =\frac{3}{2}(1+w)\;.
\end{equation}%
It is curious that $\gamma =2$ (associated with running to spectral
dimension $d_{S}=2$ in the UV) combined with radiation ($w=1/3$) in the
Einstein frame, produces de Sitter inflation in the rainbow frame. In this
case 
\begin{equation}
\tilde{t}=(\lambda k)^{2}\log t,
\end{equation}%
and so 
\begin{equation}\label{desitrain}
a\propto \exp {\left( \frac{t}{2(\lambda k)^{2}}\right) }\,,
\end{equation}%
(where the proportionality constant could be $k$-dependent).
We see that the Hubble constant is now $k$-dependent 
\begin{equation}
H=\frac{1}{2(\lambda k)^{2}},
\end{equation}%
one of the many esoteric properties of these inflationary models, and a
common feature in rainbow gravity.

\section{Intermediate inflation}

We can obtain more general inflationary solutions if we allow for more
general MDR, specifically those that in the high-energy (UV) $\lambda p\gg 1$
limit take the form: 
\begin{equation}
E^{2}\approx p^{2}g^{2}(\lambda p)\,,
\end{equation}%
where the function $g$ need not be a power-law. The speed of light is now
given by $c=E/p\approx g$, with a general profile. Just as with power-law
inflation, we can obtain intermediate inflation solutions but the
multiplicative constants will be $k$-dependent. We can
reverse engineer the MDRs associated with intermediate inflation by adapting
the argument in Section~\ref{rainbinfl}. For simplicity, let us first
illustrate the argument by the case where we start with radiation in the
Einstein frame, so that 
\begin{equation}
a(t)=a_0 t^{1/2},  \label{a1}
\end{equation}%
and seek to obtain the well-known special case of intermediate inflation with
scale-invariant inhomogeneity spectrum where $n=2/3,$ so 
\begin{equation}
a(t)\propto b(k)e^{d(k){\tilde{t}}^{2/3}},  \label{a2}
\end{equation}%
where we explicitly allow for $k$-dependence in the multiplicative factors.
From (\ref{a2}), we conclude that we must have 
\begin{equation}
{\tilde{t}}\propto (f(k)+g(k)\log a(t))^{3/2}\,,  \label{tildet2}
\end{equation}%
where $f(k)$ and $g(k)$ are still to be specified. By changing variables, we
can rewrite (\ref{tildet}) as 
\begin{equation}
{\tilde{t}}\propto (\lambda k)^{2}\int \frac{c(p)}{p^{3}}\,dp,
\label{tildet1}
\end{equation}%
where we have used (\ref{a1}) to conclude that $t\propto a^{2}$, specific to
the radiation case. By comparing (\ref{tildet1}) and (\ref{tildet2}), we can
just read off the consistency condition: 
\begin{equation}
{\tilde{t}}\propto (\lambda k)^{2}[A-\log (\lambda p)]^{3/2}\propto (\lambda
k)^{2}\int \frac{c(p)}{p^{3}}\,dp
\end{equation}%
where we have started fixing some of the free functions in the initial
ansatz. We therefore arrive at the UV-limit expression: 
\begin{equation}
c(p)\approx g(p)\propto (\lambda p)^{2}[D-\log (\lambda p)]^{1/2}\;,
\end{equation}%
where $D$ is an arbitrary ($k$-independent) constant. 
We can now check directly that this MDR results in the intermediate
inflationary solution in the dual frame: 
\begin{equation}\label{intsol1}
a\propto \lambda k\exp {\left( \frac{a_0^2 {\tilde{t}}}{2(\lambda k)^{2}}\right) }%
^{2/3}\;.
\end{equation}

This argument may be generalized to express any equation of state $w$ in the
Einstein frame as an intermediate inflationary solution in the rainbow frame
of the form $\log a\propto {\tilde{t}}^{n }$. Performing the
calculation we find that we should impose:
\begin{equation}\label{centralMDR}
E^{2}=p^{2}[1+(\lambda p)^{2\gamma}(D-\log (\lambda p))^{2\beta}],
\end{equation}%
where for completeness we have linked the IR limit with the UV solution
required. In the UV: 
\begin{equation}
g\approx (\lambda p)^{\gamma}(D-\log (\lambda p))^{\beta}\;,
\end{equation}%
and the exponents in the MDRs are related to the solutions in the Einstein
and rainbow frame by 
\begin{eqnarray}
\gamma &=&\frac{3(1+w)}{2} \\
\beta &=&\frac{1}{n }-1\,.
\end{eqnarray}%
The solution in the rainbow frame can be given more completely by: 
\begin{equation}
a\propto \lambda k\exp {\left( \frac{a_0^
\gamma\tilde{t} }{n \gamma (\lambda k)^{\gamma}}%
\right) }^{n }\;.
\end{equation}
This reduces to our illustrative solution (\ref{intsol1}) for $n=2/3$ (i.e. $\beta=1/2$) and $w=1/3$ (i.e. $\gamma=2$). It also reduces to the de Sitter case (\ref{desitrain}) for $n=1$ (i.e. $\beta =0$) and $w=1/3$ (i.e. $\gamma=2$), when the integration constants are all adjusted to be equivalent.

We note that the parameter $D$ has to be chosen so that $g$ remains positive
for a range of $p$. At face value we should conclude that $D$ 
imposes a maximal momentum, since for
\begin{equation}
p=\frac{1}{\lambda}e^D
\end{equation}
we finally get $g=0$. However we could also take the modulus and extend the 
MDRs up to infinite momentum.

\section{Fluctuations in rainbow intermediate inflation}

The conditions for viable fluctuations in models based on MDRs are entirely different from those based on standard inflation. Specifically, for $\gamma=2$ and $\beta=0$ the model is known to lead to exact scale invariance regardless of the value of $w$, within a certain range~\cite{dimred}. This corresponds to any inflationary $\tilde w$ in the rainbow frame, in contrast with standard theory (which requires near de Sitter inflation). Furthermore, within the standard theory, only intermediate inflation with the specific value $n=2/3$ leads to exact scale-invariance. Rainbow intermediate inflation may therefore be expected to predict departures from exact scale-invariance. This is confirmed by calculation.

It is easiest to perform the calculation in the Einstein frame. Then, the equation for the cosmological perturbations is:
\begin{equation}\label{veq} 
v''+\left[c^2 k^2 -\frac{a''}{a}\right]v=0, 
\end{equation}
where the prime denotes derivative with respect to conformal time $\eta$ defined with respect to the Einstein frame time (i.e. $dt=a\, d\eta$).
In terms of the variable $v$ the (comoving gauge) curvature perturbation is given by $\zeta=-v/a$. The speed of light/sound is given in the UV limit by:
\begin{equation}
c\approx (\lambda p)^{\gamma}(D-\log (\lambda p))^{\beta}\;.
\end{equation}
We want to study dual intermediate inflationary models associated with near-scale invariance. This dual requirement constrains us to select $\gamma\approx 2$ and $w\approx 1/3$. Then $a\propto \eta$ and the suitably normalized solution describing vacuum fluctuations inside the ``horizon''  is given by 
\begin{equation}\label{bc1}
v\sim\frac{e^{ik\int c d\eta}}{\sqrt{c k}}\sim 
\frac{a e^{ik\int c d\eta}}{\lambda k^{3/2} (D-\log(\lambda k/a))^{\beta/2}}\, .
\end{equation}
This should be glued to $v\sim F(k) a$ when $ck\eta\sim 1$ in order to find 
the spectrum left frozen outside the horizon (a procedure explained in 
more detail in~\cite{dimred}). To leading order the glueing point satisfies
$k (\lambda k)^2\propto \eta\propto a$, so this finally translates into:
\begin{equation}
v\sim \frac {a}{\lambda k^{3/2} (E+2 \log (\lambda k))^{\beta/2}}
\end{equation}
or
\begin{equation}\label{spec}
k^3\zeta^2\sim \frac {1}{\lambda ^2 (E+2 \log (\lambda k))^{\beta}}
\end{equation}
where $E$ is a constant. 

We see that intermediate inflation in the rainbow frame is near 
scale-invariant, with a red running spectral index. No longer is the 
$n=2/3$ case special: near-scale invariance is valid for all rainbow
intermediate inflation models, with their $n$ simply controling the power
of the logarithmic modulation.

\section{Conclusions}
The models just presented could be very interesting
observationally. The Planck satellite results have put pressure on model builders to predict departures from strict scale-invariance~\cite{Planck}. Whilst these can be easily accommodated within standard inflation, the issue arises as to how natural those departures are (or seen in another way, how predictive with respect to them the theory actually is). Intermediate inflation has long been seen as an interesting direction to explore regarding this issue~\cite{Starob,warm-inter}. In this paper we obtained intermediate inflation by postulating the appropriate MDRs in the Einstein frame capable of transforming into it in the rainbow frame. The required MDRs are of the general form (\ref{centralMDR}), which is the central result of this paper. 
The fact that we naturally obtained departures from exact scale-invariance (see Eq.~(\ref{spec})) without fine-tuning of parameters is very interesting, and will be explored in a future publication. As pointed out before (e.g. in~\cite{MagSork}) Occam's razor sometimes may dismiss the best-fit model.

It remains to understand better what MDRs of the type (\ref{centralMDR})
mean within the context of quantum gravity and the phenomenon of dimensional 
reduction in the ultraviolet. The logarithmic factor certainly has a ``renormalization'' flavour. We are currently working out the spectral dimension running function $d_S(s)$ for these MDRs, as well as an array of related implications,
with a view to clarifying their more fundamental meaning.

\section*{Acknowledgments}

We thank G. Amelino-Camelia, M. Arzano, G. Gubitosi, M. Lagos and the anonymous
referee of~\cite{rainbowred} for the discussions and comments leading to
this paper. JM was supported by an International Exchange Grant from the
Royal Society; JDB and JM both acknowledge STFC consolidated grant support.

\bibliography{refsMEAS}

\end{document}